\documentclass[%
 reprint,
 amsmath,amssymb,
 aps,
]{revtex4-2}

\usepackage{graphicx}
\usepackage{dcolumn}
\usepackage{bm}
\usepackage{braket}

\usepackage{xcolor}
\usepackage{soul}

\usepackage{makeidx}
\makeindex
\usepackage{amstext}

\begin{document}

\preprint{APS/123-QED}

\title{Constraint Analysis and Quantization of Anomalous 2-D Thomas \textendash Whitehead Gravity}

\author{Salvatore Quaid}
\email{salvatore-quaid@uiowa.edu}
\author{Vincent Rodgers}
\email{vincent-rodgers@uiowa.edu}
 
%

%
 
\affiliation{Department of Physics and Astronomy, University of Iowa, Iowa City, Iowa 52242, USA
}%

\author{Eric Biedke}
\email{biedke.1@osu.edu}

\affiliation{Department of Materials Science and Engineering, Ohio State University, Columbus, Ohio, 43210, USA}%

\date{\today}

\begin{abstract}
The two-dimensional effective Polyakov action is often realized as the anomalous contributions of string theories and fermions coupled to gravity in two-dimensions. However, as a result of the reparameterization invariance, one finds that the effective action produces vanishing Hamiltonians as constraints even in disparate gauges such as the dynamical light-cone and the ADM formalism of the metric.  On the other hand, two-dimensional gravitational theories naturally arise as geometric actions on the coadjoint orbits of the Virasoro algebra.
The Thomas \textendash Whitehead gravity formalism extends the effective Polyakov action in such a way that the defining coadjoint element for the orbit becomes a dynamical field, viz the diffeomorphism field. In this work, we examine the role the diffeomorphism field plays through the well-understood quantization of the two-dimensional anomalous contributions to gravity. This is first done in the
 dynamical light-cone and then the ADM formalisms of the metric. To examine a dynamical diffeomorphism field, the constraint analysis is then repeated in a Minkowski background. Where the dynamics of the diffeomorphism field arise from the Thomas \textendash Whitehead action. Adding dynamics to the diffeomorphism field appears to remove the vanishing Hamiltonians, however expressing the diffeomorphism in terms of the P-tensor recovers the Hamiltonian constraint. One can compare this investigation to that of a gauge Wess-Zumino-Witten action where the gauge field has become dynamical through the inclusion of a Yang-Mills term.
\end{abstract}

\maketitle


\section{\label{sec:level1}Introduction}
Originally derived by Polyakov \cite{Polyakov:1981rd,Polyakov:1987zb}, the effective Polyakov action (EPA) is the anomalous contribution of gravity in two-dimensions arising from the response of the path integral measure to Weyl and diffeomorphism transformations of the metric. Polyakov finds this action from the two-dimensional anomalous trace contribution, 
\begin{equation}
    h^{ab}\braket{T_{ab}} = \frac{d}{24\pi}R. \label{vacuum}
\end{equation}
One then asks, ``what (effective) action, would give rise to this anomalous energy-momentum tensor?''. By solving, \begin{equation}
    h^{ab}\frac{\delta F}{\delta h^{ab}} = \frac{d}{24\pi}R, \label{VaryF}
\end{equation}
one finds the effective action.  Here the metric variation of the effective action, $F$, replaces the expectation value of the energy-momentum tensor. The metric $h_{ab}$ is the two-dimensional metric with $R$ its corresponding Ricci scalar. The general solution to this is a non-local gravitational action dependent on the generic coordinates $\xi = (\sigma,\tau)$ and $\xi'= (\sigma',\tau')$ that may be written as  
\begin{multline}
    F = \frac{\mu d}{24\pi}\int{d^2\xi d^2\xi' \sqrt{h(\xi)}\sqrt{h(\xi')}R(\xi)\frac{1}{\Delta}R(\xi')}\\ + const.\int{d^2\xi \sqrt{h{(\xi)}}} \label{FAction}.
\end{multline}
The $\tfrac{1}{\Delta}$ is the Green's function for the Laplacian in the metric $h_{ab}$ when acting on a scalar. The $const.$ term is included for renormalization. It has been shown in \cite{Polyakov:1981rd,Polyakov:1987zb,Teitelboim:1983ux} that for certain choices of metric, the action is rendered local.

It is important to note that any conformal modification to Eq.[\ref{FAction}] must recover Eq.[\index{Eq}\ref{VaryF}]. However, if a non-conformal modification is made, then Eq.[\ref{VaryF}] could  also be modified. With this in mind, we recall that the effective Polyakov action can be derived by group theoretic and geometric means using the properties of the coadjoint orbits of the Virasoro algebra.    Indeed, each coadjoint orbit of the Virasoro algebra, admits a closed, non-degenerate, two-form that is governed by the quadratic differential $ \mathcal{D}_{a b}$ that transforms in the coadjoint representation of the Virasoro algebra. In the Thomas\textendash Whitehead formalism \cite{Brensinger:2017gtb,Brensinger:2020gcv}, $ \mathcal{D}_{a b}$  is treated as a fundamental field that arises from a projective connection.   $\mathcal{D}_{a b}$ (as well as just $\mathcal{D}$) is called the diffeomorphism field. The isotropy algebra of  $\mathcal{D}$  is    defined by those vectors fields,  $\xi,$  where their Lie derivative kills the diffeomorphism field $\mathcal{D}$, i.e.  $\mathcal{L}_{\xi} \mathcal{D}=0.$  These vector fields generate the divisor group, $\mathcal{H}$  of the coset space $\text{diff}(S_1)/\mathcal{H}$, where $\text{diff}(S_1)$ defines the Virasoro group.   This was demonstrated in \cite{Rai:1989js,Alekseev:1988ce,Alekseev:1988vx,Delius:1990pt}.  In the next section, we will show that the presence of the diffeomorphism field modifies Eq.[\ref{FAction}].

In what follows the EPA is  investigated for two choices of the metric, namely the light-cone gauge and the ADM formalism. Written in Polyakov's dynamical light-cone gauge, the metric takes the form,
\begin{equation}
    h_{ab} = \begin{pmatrix}
        0 && 1/2\\
        1/2 && h_{++}
    \end{pmatrix}, \;\; h_{++} = \frac{\partial_+ f}{\partial_- f}.
\end{equation}
In this gauge, $x^+$ is to be treated as the dynamical coordinate and $f(x^+,x^-)$ is a complex function associated with the residual symmetry from gauge-fixing. This becomes dynamical through the EPA action as 
\begin{equation}
    F = \frac{\mu d}{24\pi}\int{\left(\tfrac{(\partial_-\partial_+ f)(\partial_-^{\;2}f)}{(\partial_- f)^2} - \tfrac{(\partial_+ f)(\partial_-^{\;2}f)^2}{(\partial_-f)^3}
    \right)}d^2x.
\end{equation}

 In the ADM formulation, as in \cite{Teitelboim:1983ux}, the two-dimensional metric is taken to be
\begin{equation}
     h_{ab} = e^\varphi \begin{pmatrix}
        (\eta^1)^2-(\eta^\perp)^2 && \eta^1\\
        \eta^1 && 1
    \end{pmatrix},
\end{equation}
where $\eta^\perp$ is the lapse function, $\eta^1$ is the shift function, and together $\eta^a = (\eta^\perp,\eta^1)$ form the time vector field. The Weyl anomaly is now recovered as
\begin{equation}
    \frac{\delta F}{\delta \varphi}
 = h^{ab}\frac{\delta F}{\delta h^{ab}} = -\frac{1}{\kappa}\sqrt{g}(R-\Lambda_0).
 \end{equation}
 This form of the  anomaly offers a clearer connection to the canonical quantization of general relativity. The effective action now takes the form
 \begin{multline}
            F = \tfrac{1}{2\kappa}\int{dxdt \left(\tfrac{1}{\eta^{\perp}}(\dot{\varphi} - \varphi'\eta^1-2(\eta^1)')^2 - \eta^\perp (\varphi')^2 \right.}\\
            \left.+ 4\eta^\perp \varphi'' + 2\eta^\perp \Lambda_0 e^{\varphi} 
            \right). 
\end{multline}
The scalar field $\varphi$ is treated as a generic metric scaling. The time vector field in the ADM formalism enjoys $\eta^a\nabla_a t = -1$. Within the ADM formalism the often investigated conformal metric can be seen as the proper-time gauge \cite{TeitelboimPTG}. A comparison of the EPA in the dynamical light-cone versus that of the Euclidean conformal gauge can be found in \cite{Polyakov:1981rd,DANIELSSON1989292}.

These actions will be modified by the inclusion of the diffeomorphism field. The case of a background diffeomorphism field will be shown to contribute a traceless term to the metric field equation. This approach of modifying the effective Polyakov action differs from others as the diffeomorphism field arises from the geometry and is not motivated by matter. That is the diffeomorphism field is to be taken as separate from the stress-energy tensor.

\subsection{Geometric Actions}
Shown by \cite{Alekseev:1988ce,Rai:1989js} the EPA can be derived by the construction of the geometric action for the Virasoro algebra. Similarly, the Wess\(-\)Zumino\(-\)Witten Effective action (WZW) can be found as the geometric action for the Kac-Moody algebra.

Elements of the Kac-Moody algebra produce the commutation law
\begin{equation}
    [J^{\alpha}_n,J^{\beta}_m] = if^{\alpha \beta \gamma}J^{\gamma}_{n+m} + nk\delta_{n+m,0}\delta^{\alpha\beta},
\end{equation}
The geometric action produced finds the partially gauge-fixed Wess\(-\)Zumino\(-\)Witten action
\begin{multline}
    S = \frac{-k}{2\pi}\int{d^2\sigma Tr(Ag^{-1}\partial_\tau g)}\\ - \frac{\mu k}{4\pi}\int{d^2\sigma Tr(g^{-1}\partial_\sigma g g^{-1} \partial_\tau g)}\\
    +\frac{\mu k}{4\pi}\int{d^3\sigma Tr(g^{-1}\partial_\sigma g[g^{-1}\partial_\lambda g, g^{-1}\partial_\tau g])}.
\end{multline}
Likewise, elements of the Virasoro algebra produce
\begin{equation}
    [L_m, L_n] = (m-n)L_{m+n} + am^3\delta_{m+n,0},
\end{equation}
While the geometric action is found to be
\begin{multline}
   F = \frac{\mu d}{24\pi}\int{\left(\frac{(\partial_\sigma \partial_\tau f)(\partial_\sigma^{\;2}f)}{(\partial_\sigma f)^2} - \frac{(\partial_\tau f)(\partial_\sigma^{\;2}f)^2}{(\partial_\sigma f)^3}
    \right)}d^2\sigma\\
    + \int{\mathcal{D}\frac{\partial_\tau f}{\partial_\sigma f}}d^2\sigma.
    \label{EPAR&R}
\end{multline}
Compared to the gauge-fixed WZW action, one can see it is premature to restrict the new field, $\mathcal{D}$, to be a constant. One could further generalize the theory through the inclusion of a \textit{bare} cosmological constant that takes the form originally included by Polyakov. Furthermore, as this new contribution transforms in the coadjoint representation, one can observe its role in defining which orbit the theory lives on. Choosing this field to be a constant, as has been done by previous quantization attempts, investigates only a single orbit. Integration over the diffeomorphism field allows one to move across these orbits. Allowing for the field to be dynamical therefore allows an understanding of the generated symmetries. The role of the diffeomorphism field can be compared to that of the gauge field in the Kac-Moody algebra. This problem is well known in string theory, \cite{Witten:1987ty}. However, its importance to the quantization of gravitational theories such as general relativity is also apparent, as general relativity contains the same symmetries (in two-dimensions) and thus the same effective theory.

Recent work in projective gravitational theories by \cite{Rai:1989js, Brensinger:2020gcv, Rodgers:2006ep} shows this term to be related to a portion of the projective connection known as the diffeomorphism field. As the diffeomorphism field is taken as a piece of a connection written without torsion, we take $\mathcal{D}_{ab}$ to be symmetric. Specifically, in the two-dimensional light-cone gauge, one finds $\mathcal{D}$ to be the $\mathcal{D}_{--}$ piece of the connection. The generalized version of Eq.[\ref{EPAR&R}] is then 
\begin{multline}
   F = \frac{\mu d}{24\pi}\int{d^2\xi d^2\xi' \sqrt{h(\xi)}\sqrt{h(\xi')}R(\xi)\frac{1}{\Delta}R(\xi')}\\
    + \int{d^2\xi\sqrt{h(\xi)}\mathcal{D}_{ab}(\xi)h^{ab}(\xi)}.
\end{multline}
One recovers the EPA in the light-cone gauge by taking the $\mathcal{D}_{-+}$ term to be constant and vanishing on the boundary.

The coupling of the EPA to the background diffeomorphism field will be shown to recover Eq.[\ref{VaryF}]. Making the diffeomorphism field dynamical  will modify Eq.[\ref{VaryF}].
The diffeomorphism fields' contribution to the EPA will now be shown not to violate the trace anomaly. Defining the EPA as $F = F_1 + F_2$, where $F_2$ contains the diffeomorphism contribution, recovering the trace anomaly requires
\begin{equation}
    h^{ab}\frac{\delta F_2}{\delta h^{ab}} = 0.
\end{equation}
From direct calculation, we find
\begin{equation}
    h^{ab}\frac{\delta F_2}{\delta h^{ab}} = \frac{d-2}{2}\sqrt{h}\mathcal{D}
\end{equation}
Where we have defined $\mathcal{D}\equiv h^{ab}\mathcal{D}_{ab}$. It is thus straightforward to see that for two-dimensional theories, the diffeomorphism field's contribution to the Weyl anomaly vanishes. Including the dynamics of the diffeomorphism field does not recover this form of the trace anomaly, but instead an appropriately modified form. This allows us to see the diffeomorphism field as the natural generalization of the effective theories. It can also be seen from the form of $F_2$ that, dependent on the metric gauge, the diffeomorphism field can have a non-trivial influence on the constraint structure.

The dynamics for the diffeomorphism field arises from Thomas \textendash Whitehead gravity (TW gravity) \cite{Brensinger:2017gtb,Brensinger:2020gcv}.   TW gravity has its origins as Lagrangian on the  Thomas \textendash \ Whitehead bundle.  The Thomas \textendash \ Whitehead bundle  is a line bundle over the space of equivalence classes of all  Riemannian connections that are projectively equivalent. Two connections are projectively equivalent when \[\hat\Gamma^{a}_{\;\;bc}=\Gamma^{a}_{\;\;bc} + \delta^a_{\;b}\omega_c + \delta^a_{\;c} \omega_{b}, \] for any one-form $\omega$.   This line bundle is called a Thomas Cone \cite{BaileyT.N.1994TSBf,CapA.2014Emip,Crampin,curry_gover_2018,Eastwood,Eastwood2,GoverA.Rod2012DEgE} and is  equipped with its own connection. Part of this connection contains the projective invariant \cite{Thomas:1925a,Thomas:1925b,Whitehead}
$\Pi^{a}_{\;\;bc}$, which can be constructed from any member of the equivalence class \( |\Gamma^{a}_{\;\;bc}|\) as,
\begin{equation}
    \Pi^{a}_{\;\;bc} = \Gamma^{a}_{\;\;bc} + \delta^a_{\;b}\alpha_c + \delta^a_{\;c}\alpha_{b}.
\end{equation}
Here  $\alpha_a$ is the trace of  connection,
\begin{equation}
    \alpha_a = \frac{-1}{d+1}\Gamma^e_{\;\;ea}.
\end{equation}
From here one can construct the Thomas projective invariant curvature  pseudo-tensor $\mathcal{R}^a_{\;\;bcd}$, 
\begin{equation}
    \mathcal{R}^a_{\;\;bcd} = \partial_c \Pi^a_{\;\;db} - \partial_d \Pi^a_{cb} + \Pi^{a}_{\;\;ce}\Pi^{e}_{\;\;db}-\Pi^{a}_{\;\;de}\Pi^{e}_{\;\;cb}, 
\end{equation} 
an its associated projective invariant Ricci pseudo-tensor.  
In TW Gravity the projective invariant Ricci symbol is replaced with the diffeomorphism field,  $\mathcal{D}_{ab}$ \cite{Brensinger:2017gtb,Brensinger:2020gcv}.  The diffeomorphism field then gets dynamics through a ``curvature squared'' Lagrangian.  In this way $\mathcal{D}_{ab}$ is independent of the metric and in the one-dimensional limit, transforms as a coadjoint element of the Virasoro algebra making contact with string theory.      The  Thomas \textendash Whitehead gravitational Lagrangian is constructed to collapse to the Einstein-Hilbert action in four-dimensions  when $\mathcal{D}_{ab} $ vanishes and when the connection $\Gamma^{a}_{\;\;bc}$   is compatible with the metric.

To proceed, we will first investigate the constraint analysis with the diffeomorphism field as only a background field.  Later, we will include the dynamical contributions for the diffeomorphism field but keeping the metric as a background field.

Explicitly, the dynamics for the diffeomorphism field come from the projective Gauss \(-\) Bonnet (PGB) terms in the Thomas \textendash Whitehead gravity action.  We write this as 
\begin{multline}
    S_{PGB} = J_0c\lambda_0^2\int{d^dx\sqrt{|h|}\;K_{bcd}K^{bcd}}\\
    -J_0c\int{d^dx\sqrt{|h|}\;(\mathcal{K}^a_{\;\;bcd}\mathcal{K}_{a}^{\;\;bcd}-4\mathcal{K}_{ab}\mathcal{K}^{ab}+\mathcal{K}^2}),
    \label{PGBaction}
\end{multline}
where the $\mathcal{K}$-curvatures,  $\mathcal{K}^\mu_{\;\;\alpha \beta \nu}$, have non-vanishing terms for the space-time components, 
\begin{equation}
    \mathcal{K}^a_{\;\; bcd} = \mathcal{R}^a_{\;\;bcd} + \delta^a_{\;\;c}\mathcal{D}_{db}-\delta^a_{\;\;d}\mathcal{D}_{cb}
\end{equation}
\begin{equation}
    \lambda  \breve{\mathcal{K}}^\lambda_{\;\;bcd}=\mathcal{K}^\lambda_{\;\;bcd} =\lambda \partial_{[a}\mathcal{D}_{b]c}+\lambda \Pi^d_{\;\;c[b}\mathcal{D}_{a]d}.
\end{equation}
In the projective Gauss\(-\)Bonnet action, Greek indices refer to the projective space, while Latin indices are taken to be space-time. $\lambda$ is the projective fibre coordinate, and $\lambda$ as an index that is reserved for the projective direction.

The projective Cotton\(-\)York terms appearing in the projective Gauss\(-\)Bonnet action are found as
\begin{equation}
    K_{bcd} = g_a \mathcal{K}^a_{\;\; bcd} + \breve{\mathcal{K}}^\lambda_{\;\;bcd},
\end{equation}
where pseudo one-form $g_a$ is defined as
\begin{equation}
    g_a =  \tfrac{-1}{d+1} \partial_a \log(\sqrt{h}).
\end{equation}
Metrics written in constant volume coordinates yield $g_a = 0$, and similarly, traceless connections enjoy $\alpha_a = 0$. The dynamical light-cone metric can be seen to satisfy these conditions, while the ADM formalism does not. Again, for the metric-compatible connection and taking the diffeomorphism field to vanish recovers the Gauss\(-\)Bonnet action from the projective Gauss\(-\)Bonnet action. Furthermore, the projective Ricci scalar for a metric-compatible connection is simply
\begin{equation}
    \mathcal{K} = \mathcal{R} + (d-1)\mathcal{D}.
\end{equation}
This demonstrates the recovery of the Einstein-Hilbert action in the limit $\mathcal{D}_{ab}\rightarrow 0$ for two and four dimensions.

Although the Gauss\(-\)Bonnet action vanishes identically in two-dimensions, its projective counterpart will not.  This non-vanishing allows the investigation into a dynamical diffeomorphism field in two-dimensions. However, as the Gauss\(-\)Bonnet action produces no derivatives of the metric higher than second order, the presence of the derivatives in the dynamical light-cone metric can ultimately produce third-order derivatives. This is the primary motivation for the abandonment of this metric in favor of the ADM formalism. The lack of $\alpha_a = g_a =0$ in the ADM formalism will greatly complicate the constraint structure. Early work on the quantization and constraint analysis of the diffeomorphism field as a potential squared theory inspired from the WZW gauge field can be found in \cite{Lano:1994gx}. The foundations of the canonical quantization and constraint analysis explored in this paper can be found in \cite{Sal}. 

Attempting to investigate the fully dynamical theory often encounters difficulties in the form of higher derivatives. These can be avoided by trading the diffeomorphism field as the canonical variable for the P-tensor found in \cite{Brensinger:2020gcv}. Utilizing the one-form, $\alpha_a$, we are able to construct an object that transforms as a tensor, $\mathcal{P}_{ab}$. This tensor takes the form
\begin{equation}
    \mathcal{P}_{ab} = \mathcal{D}_{ab}-\partial_a \alpha_b + \Gamma^{c}_{\;\;ab}\alpha_c + \alpha_a \alpha_b.
    \label{ptensor}
\end{equation}
In this formalism the projective Ricci scalar becomes
 \begin{equation}
     \mathcal{K} = R + (d-1)\mathcal{P}.
 \end{equation}
Where $R$ is the Ricci scalar constructed from the metric-compatible space-time connection and $\mathcal{P}$ is the space-time metric trace of $\mathcal{P}_{ab}$.

\section{Background Diffeomorphism Field}

\subsection{Dynamical Light-Cone Gauge}
The geometric action recovers the effective Polyakov action written in the light-cone gauge. This is also the form of the effective action favored by Polyakov in \cite{Polyakov:1987zb} and demonstrates the relationship between the effective action and string theory the clearest. In the dynamical light cone gauge, the action is 
\begin{multline}
    F = \frac{\mu d}{24\pi}\int{\left(\frac{(\partial_- \partial_+ f)(\partial_-^{\;2}f)}{(\partial_- f)^2} - \frac{(\partial_+ f)(\partial_-^{\;2}f)^2}{(\partial_- f)^3}
    \right)}d^2 x\\
    + \frac{1}{2}\int{\mathcal{D}_{--}\frac{\partial_+ f}{\partial_- f}}d^2 x.
\end{multline}

Although the action does not have Weyl invariance, a $U(1)$ scaling symmetry of the metric field $f(x^{+},x^{-})$ is apparent in the metric and leaves the action invariant. From this, the currents $j_+$ and $j_-$ can be found.
We begin by performing integration by parts and taking the boundary terms to vanish. The action can then be written as linear in $\partial_+ f$ and takes the form
\begin{multline}
    F = \frac{\mu d}{24\pi}\int{\left(\frac{(\partial_-^{\;2} f)^2}{(\partial_- f)^3} - \frac{(\partial_-^{\;3}f)}{(\partial_- f)^2}
    \right)(\partial_+f)}d^2x\\
    +\frac{1}{2}\int{\mathcal{D}_{--}\frac{\partial_+f}{\partial_- f} d^2x}.
\end{multline}
The currents are defined as
\begin{equation}
    j^+ \equiv \frac{\delta F}{\delta (\partial_+ f)},\;\text{and } j^- \equiv \frac{\delta F}{\delta (\partial_- f)}
\end{equation}
These are straightforward to read off as
\begin{equation}
    j^+ = \frac{\mu d}{24\pi}\left(\frac{(\partial_-^{\;2} f)^2}{(\partial_- f)^3} - \frac{(\partial_-^{\;3}f)}{(\partial_- f)^2}
    \right) + \frac{1}{2}\frac{\mathcal{D}_{--}}{\partial_- f}
\end{equation}
and
\begin{multline}
    j^- = \frac{\mu d}{24\pi}\left(\frac{2(\partial_-\partial_+ f)(\partial_-^{\;2} f)}{(\partial_- f)^3}-\frac{3(\partial_+ f)(\partial_-^{\;2} f)^2}{(\partial_- f)^4}\right.\\ - \left. \frac{(\partial_-^{\;2}\partial_+ f}{(\partial_- f)^2} + \frac{2(\partial_+ f)(\partial_- ^{\;3} f)}{(\partial_- f)^3}
    \right)\\
    -\frac{1}{2}D_{--}\frac{(\partial_+f)}{(\partial_- f)^2}.
\end{multline}

Defining now $j^\mu = (j^-,j^+)$, one can show that in general, the divergence of the current for this symmetry recovers
\begin{equation}
    \partial_\mu j^\mu = \frac{\delta F}{\delta f}.
\end{equation}
As the Christoffel connections found from the dynamical light-cone metric are traceless, the previous statement is covariant. The right-hand side of the divergence statement is the field equation with respect to the scalar field $f$. Thus, the divergence does not vanish trivially, but from enforcing the field equation to be on shell. This symmetry refers to the conservation of scale, not in the metric but in the dynamical field that the metric itself depends on.
Utilizing the currents, we can write the action simply as
\begin{equation}
    F = \int{d^2x\; j^+ (\partial_+f)}
\end{equation}

Seeking to canonically quantize the theory, we construct the canonical momenta to $f$ as
\begin{equation}
    p \equiv \frac{\delta F}{\delta (\partial_+ f)}.
\end{equation}
This can be seen to be equal to $j^+$. As the current does not depend on $(\partial_+f)$, we promote this canonical momenta to the primary constraint as
\begin{equation}
    \phi \equiv p - j^+ \approx 0.
\end{equation}
Following Dirac's constraint procedure, $\approx \rightarrow =$ on the constraint surface, we find Hamilton's equations of motion through the promotion of Poisson brackets to Dirac brackets.

Attempting to construct a Hamiltonian density for this theory results in
\begin{equation}
    \mathcal{H} = V(x^-,x^+)\phi.
\end{equation}
Where $V(x^-,x^+)$ is a Lagrange multiplier that can be identified with $(\partial_+ f)$. As $\phi$ is our primary constraint, so is the Hamiltonian density, and as $\phi \approx 0$, so too does $\mathcal{H} \approx 0$. This Hamiltonian constraint is similar to that of the Polyakov action, which generates a Virasoro algebra.

Attempting to search for secondary constraints, we define $\chi \equiv \{\phi, H\} \approx 0$. The resulting expression for $\chi$ is long and, as such, will not be written here. However, $\chi$ can easily be seen to be a functional of the Lagrange multiplier and thus is not to be treated as a constraint but as a consistency condition.

As the only constraint present in the theory is the primary constraint $\phi$, it is trivial to show $\phi$ is first class. As a direct consequence of this, the derived Dirac brackets recover the Poisson brackets. Utilizing the Dirac brackets, we find the dynamics of the phase-space variables. However, as the constrained phase-space of this theory contains a single field, Hamilton's equations do not reveal new information. That is, the dynamics of $f$ can be found as
\begin{equation}
    \frac{\partial f}{\partial x^+} = V(x^-,x^+)
\end{equation}
This confirms our earlier identification of $(\partial_+ f)$ with the Lagrange multiplier $V(x^-,x^+)$. A proper investigation into what physics the Lagrange multiplier imposes on the larger system will not be done here.

The equal-time Dirac brackets of the phase-space variables amongst themselves reveal
\begin{multline}
    \{f(x^-,x^+),f(y^-,x^+)\} = 0,\\ \{p(x^-,x^+),p(y^-,x^+)\} = 0,\\
    \{f(x^-,x^+),p(y^-,x^+)\} = \delta(x^- - y^-).
\end{multline}
From such, the equal time canonical commutation relations are simply
\begin{multline}
    [\hat{f}(x^-,x^+),\hat{f}(y^-,x^+)] = 0,\\ [\hat{p}(x^-,x^+),\hat{p}(y^-,x^+)] = 0,\\
    [\hat{f}(x^-,x^+),\hat{p}(y^-,x^+)] = i \delta(x^- - y^-).
\end{multline}
Written in the position representation, we promote the fields to operators as
\begin{multline}
    f(x^-,x^+) \rightarrow \hat{f}(x^-,x^+) = f(x^-,x^+),\\  p(x^-,x^+) \rightarrow \hat{p}(x^-,x^+) = -i\frac{24\pi}{\mu d}\frac{\delta}{\delta f(x^-,x^+)}.
\end{multline}

Acting on a state $\ket{\psi}$, the Hamiltonian can be written as
\begin{equation}
    \hat{H}\ket{\psi} \approx 0 \rightarrow \hat{\phi}\ket{\psi} \approx 0.
\end{equation}
In the position representation, the Hamiltonian constraint takes the form of an Einstein-Schrodinger equation:

\begin{equation}
    \frac{24\pi}{\mu d}\frac{\delta \Psi}{\delta f} -i  \hat{j}^{+} \Psi \approx 0.
\end{equation}
The extension of $\hat{j}^{+}$ from $j^{+}$ is trivial in this representation, as $j^{+}$ is independent of $p(x^-,x^+)$. The $\Psi$ variable is a general wave functional of the scalar field $f(x^-,x^+)$.

Exact solutions of the equation cannot be found until specific forms of $j^{+}$ are chosen. However, as $j^{+}$ depends on both $f(x^-,x^+)$ and $\mathcal{D}_{--}(x^-,x^+)$, choosing $j^{+}$ amounts to choosing forms of the metric field and the diffeomorphism field. As both these fields are comments on the underlying geometry of the problem, exact solutions to the Hamiltonian cannot be found until exact forms of the geometry are specified. However, a generic solution to the Hamiltonian constraint may be written as
\begin{equation}
    \Psi = e^{i\tfrac{\mu d}{24 \pi}A[f(x^-,x^+)]}
\end{equation}
Where the functional $A[f]$ is found from $j^+$ as
\begin{equation}
    j^+ = \frac{\delta A}{\delta f}.
\end{equation}
    
Although this second quantization of the theory is under-defined, we can proceed with the classical theory by solving the constraint in favor of $\mathcal{D}_{--}$. That is
\begin{equation}
    \mathcal{D}_{--} = 2\left((\partial_- f)p- \frac{\mu d}{24\pi}\frac{(\partial_-^{\;3}f)}{(\partial_- f)} + \frac{\mu d}{24\pi} \frac{(\partial_-^{\;2}f)^2}{(\partial_- f)^2} \right).
    \label{Dtof1}
\end{equation}

For a given background diffeomorphism field Eq.[\ref{Dtof1}] determines the form of the metric. Writing $f$ and $p$ in terms of a Fourier mode expansion, such that the commutation relations are satisfied, allows us to write the diffeomorphism field component in terms of quantum operators. A generic mode expansion of $f$ and $p$ is given by
\begin{equation}
    f = \int{\frac{dk}{2\pi}a_k e^{i(kx^-+\omega x^+)}}
\end{equation}
and
\begin{equation}
    p = \int{\frac{dk'}{2\pi}i a_{k'}^{\dagger}e^{-i(k'x^-+\omega x^+)}}.
\end{equation}
Where we impose $[\hat{a}_k,\hat{a}^\dagger_{k'} ]=\delta_{k k'}$ and $[\hat{a}_k,\hat{a}_{k'} ]=[\hat{a}^\dagger_{k},\hat{a}^\dagger_{k'} ]=0$. It can be seen that the second and third terms in Eq.[\ref{Dtof1}] are the contributions from the original EPA, and that for such a mode expansion, cancel. The surviving terms find
\begin{equation}
        \mathcal{D}_{--}(x^-) = -\int{\frac{dk}{\pi}\frac{dk'}{2\pi}k (a_k a_{k'}^{\dagger} + c) e^{ix^-(k-k') }}
\end{equation}
Where $c$ is an unknown operator ordering ambiguity constant. This describes the necessary Fourier expansion of the diffeomorphism field on the constraint surface. The presence of the exponential in terms of $x^-(k-k')$ inspires us to integrate over $\mathcal{D}_{--}$ in favor of a Dirac delta. We define the new form of the diffeomorphism in which the $x^-$ dependence has been integrated out as
\begin{equation}
    D_{--} \equiv \int_{-\infty}^{\infty}{dx^- \mathcal{D}_{--}}.
\end{equation}
For which we find
\begin{equation}
        D_{--} = -\frac{1}{\pi}\int{dk\;k (a_k a_{k}^{\dagger} + c)}.
\end{equation}
This can be identified as the number operator with an extra $k$ dependence.

 The integration bounds of $dk$ have been left ambiguous, as integrating over all $k$ produces an infinite diffeomorphism field. To recover a non-infinite contribution upper/lower cutoffs must be included. In section $III$ the dynamics of the diffeomorphism are investigated with a Minkowski metric. For the case where the background diffeomorphism field is ruled by this simple dynamical diffeomorphism theory the gravitational wave dispersion relation and speed of light solutions provide a natural choice of cut off. The limits of the integration become
\begin{equation}
    k_a = \frac{i\sqrt{2}}{\lambda_0},\;\; k_b = \frac{\sqrt{{-2+\omega^2 \lambda_0^2}}}{\lambda_0}.
\end{equation}
From the number operator component of the diffeomorphism field, it can be seen that all transition amplitudes will vanish. Defining general transition amplitudes/expectation values as  $ \braket{D}  \equiv \bra{m}D\ket{n}$ finds
\begin{equation}
    \braket{D_{--}}  = \frac{-1}{\pi}\int^{k_b}_{k_a}dk\; k((n +1 +c)\delta_{mn}
\end{equation}
Carrying out the remaining integration utilizing the background metric solutions reveals
\begin{equation}
     \braket{D_{--}}   = -\frac{\omega^2}{2\pi}
    (n +1 +c)\delta_{mn}.
\end{equation}

An alternative to the gravitational dispersion relation, a speed of light solution with $k = \omega$ can also be found. Changing the integration bounds to be $k_a = 0$ and $k_b =\omega$ again finds $ \braket{D_{--}} $ to be
\begin{equation}
     \braket{D_{--}}   =-\frac{\omega^2}{2\pi}
    (n +1 +c)\delta_{mn}.
\end{equation}
Both wave propagation solutions result in the same form for the quantized diffeomorphism field. Thus the expectation value for the diffeomorphism field defines the quantum state of the metric.

\subsection{The ADM Metric}
Choosing a metric gauge that contains no derivatives of the phase-space fields while producing a local effective action naturally leads to the ADM formalism. The effective action with a background diffeomorphism field is given by
 \begin{multline}
            F = \tfrac{1}{2\kappa}\int{d^2x \left(\tfrac{1}{\eta^{\perp}}(\dot{\varphi} - \phi'\eta^1-2(\eta^1)')^2 - \eta^\perp (\varphi')^2 \right.}\\
            \left.+ 4\eta^\perp \varphi'' + 2\eta^\perp \Lambda_0 e^{\varphi} + \tfrac{1}{\eta^\perp}((\eta^{\perp})^2\mathcal{D}_{11}\right.\\\left.
            -(\eta^{1})^2\mathcal{D}_{11} + 2\eta^1 \mathcal{D}_{10}-\mathcal{D}_{00}
            \right) 
            \label{ADMEPAaction}
\end{multline}
As the inclusion of the diffeomorphism field produces terms mixing the lapse and shift functions, it can already be seen that the Hamiltonian constraint and the time vector fields' role as a Lagrange multiplier is complicated. The $U(1)$ symmetry enjoyed in the dynamical light-cone gauge is no longer obviously manifested.

The canonical momenta are found as
\begin{equation}
    p \equiv \frac{\delta F}{\delta \dot{\varphi}} = \frac{1}{2\kappa \eta^\perp}\left(2\dot{\varphi}-2\varphi'\eta^1 -4(\eta^1)'\right),
\end{equation}
and
\begin{equation}
    p_\perp  \equiv \frac{\delta F}{\delta \dot{\eta}^\perp} = 0,\;\; p_1  \equiv \frac{\delta F}{\delta \dot{\eta}^1} = 0
\end{equation}
The primary constraints arise from the lapse and shift functions corresponding momenta and are defined as
\begin{equation}
            \mathcal{\phi}_\perp \equiv p_\perp\approx 0,\text{and} \;\;\mathcal{\phi}_1 \equiv p_1\approx 0.
\end{equation}
These primary constraints with a naive Hamiltonian can be used to produce the secondary constraints
\begin{equation}
        \chi_1 = -\left(p\varphi'-2p'+2\frac{\eta^1}{\eta^\perp}\mathcal{D}_{11} - 2\frac{1}{\eta^\perp}\mathcal{D}_{10}.
        \right),
\end{equation}
and
\begin{multline}
        \chi_\perp =-\left(\frac{\kappa}{2}p^2 - \frac{2}{\kappa}\varphi'' +\frac{1}{2\kappa}(\varphi')^2 - \frac{\Lambda_0}{\kappa}e^{\varphi}
        \right.\\\left.-\mathcal{D}_{11}-\frac{(\eta^1)^2}{(\eta^\perp)^2}\mathcal{D}_{11}+2\frac{\eta^1}{(\eta^\perp)^2}\mathcal{D}_{10}-\frac{1}{(\eta^\perp)^2}\mathcal{D}_{00}
        \right).
\end{multline}
All of which can be shown to be second-class and produce the constraint matrix

\begin{widetext}
\begin{equation}
    \mathcal{M} = \begin{pmatrix}
        0 && 0 && \frac{2}{\eta^\perp}\mathcal{D}_{11} && \frac{2}{(\eta^\perp)^2}(\mathcal{D}_{10}-\eta^1\mathcal{D}_{11})\\
        0 && 0 && \frac{2}{(\eta^\perp)^2}(\mathcal{D}_{10}-\eta^1\mathcal{D}_{11})&& \frac{2}{(\eta^\perp)^3}((\eta^1)^2\mathcal{D}_{11}-2\eta^1\mathcal{D}_{10}+\mathcal{D}_{00})\\
        - && - && 0 && \frac{\Lambda}{\kappa}\varphi' e^\varphi - \kappa p'p - \frac{1}{\kappa}\varphi' \varphi''\\
        - && - && -&& 0
    \end{pmatrix}\delta(x-x').
    \label{eq:wideeq}
\end{equation}
\end{widetext}
The $-$ terms can be inferred from the antisymmetric properties of the Poisson brackets. 

Using the Dirac brackets to source Hamilton's equations of motion, we are now free to fully constrain the Hamiltonian. The full form of the Hamiltonian density is given by
\begin{multline}
        \mathcal{H}_2 = -\eta^{\perp}\chi_\perp - \eta^1\chi_1 + \frac{2}{\eta^\perp}\left(\mathcal{D}_{00} - \eta^1\mathcal{D}_{10}
        \right)\\
        +\phi_1v_1 + \phi_\perp v_\perp.
\end{multline}
Thus, it has been shown the presence of the diffeomorphism field no longer allows $\eta^\perp$ and $\eta^1$ to be identified purely as the Lagrange multipliers. Constraining the Hamiltonian density produces
\begin{equation}
        \mathcal{H}_c =  \frac{2}{\eta^\perp}\left(\mathcal{D}_{00} - \eta^1\mathcal{D}_{10}
        \right).
        \label{CHADM}
\end{equation}
The limit $\mathcal{D}_{ab} \rightarrow 0$ recovers both the Hamiltonian constraint and secondary constraints from the original theory. The Hamiltonian now takes on a non-trivial form that appears to allow evolution of states. However, it will be shown that consistency conditions arise, further constraining the Hamiltonian. The constrained Hamiltonian, Eq.[\ref{CHADM}], and the derived Dirac brackets finds the equations of motion
\begin{equation}
   \dot{\varphi} = 0,
\end{equation}
\begin{equation}
    \dot{\eta}^\perp = \frac{-\mathcal{D}_{10}(\eta^\perp)^3}{2\kappa Det(\mathcal{D})}(\kappa^2 pp' -\varphi' (e^\varphi \Lambda_0 -\varphi'')),
\end{equation}

\begin{multline}
    \dot{\eta}^1 = \frac{(\eta^\perp)^2}{2\kappa Det(\mathcal{D})}(\mathcal{D}_{00}-\mathcal{D}_{10}\eta^1)(\kappa^2 p p' \\-\varphi'(e^\varphi \Lambda_0 - \varphi'')),
\end{multline}
\begin{equation}
    \dot{p} = 0,
\end{equation}
\begin{equation}
    \dot{p}_\perp = \frac{2}{(\eta^\perp)^2}(\mathcal{D}_{00} - \mathcal{D}_{10}\eta^1),
\end{equation}
and
\begin{equation}
    \dot{p}_1 = \frac{2}{\eta^\perp}\mathcal{D}_{10}.
\end{equation}
The dynamics of $\varphi$ and $p$ demonstrate a time independence in the metric scaling field and its canonical momenta. Enforcing the primary constraints in the last two equations of motion reveals constancy conditions on the form of the diffeomorphism field. 
Further constraining the Hamiltonian using the equations of motion recovers $\mathcal{H}_c \approx 0$.

The Dirac brackets also find the commutation relations of the phase-space fields
\begin{multline}
    \{\eta^\perp,p_\perp\} = \delta(x-x'), \{\eta^1,p_1\} = \delta(x-x'),\\ \text{ and } \{\varphi,p\} = \delta(x-x').
\end{multline}
These are the expected canonical relations. However, we now gain the new non-trivial commutation relation
\begin{equation}
    \{\eta^1,\eta^\perp\} = \frac{\delta(x-x')(\eta^\perp)^4}{4\kappa Det(\mathcal{D})}(\kappa^2 pp' -\varphi'(e^\varphi \Lambda_0 -\varphi'')).
\end{equation}
All other combinations of phase-space fields have zero Dirac brackets.

As the theory contains a non-trivial dependence on the lapse-shift function, investigating minisuperspace restrictions of the metric must be done with caution. Unfortunately, as the ADM formalism assumes asymptotic flatness, whereas the Friedmann-Robertson-Walker metric for two-dimensions recovers a two-sphere, the application to cosmological models is not obvious and will be the focus of future work. However, the reduction to the conformal metric via the proper-time gauge can still be done.
\subsubsection{The Proper-Time Gauge}
This gauge is realized as $\eta^\perp \rightarrow const.$ and $\eta^1 \rightarrow 0$. We will further choose $\eta^\perp \rightarrow 1$ for simplicity. This gauge fixing is to be done prior to reducing to the constraint surface and produces the Hamiltonian density
\begin{equation}
    \mathcal{H}_c = 2\mathcal{D}_{00},
\end{equation}
the secondary constraints become
\begin{equation}
         p\varphi' - 2p' - 2\mathcal{D}_{10} \approx 0,
\end{equation}
and
\begin{equation}
        \frac{\kappa}{2}p^2 + \frac{1}{\kappa}\varphi'' -\frac{1}{2\kappa}\varphi'^2 + \frac{\Lambda_{0}}{\kappa}e^\varphi - \mathcal{D}_{11} - \mathcal{D}_{00} \approx 0.
\end{equation}
While the equations of motion become
\begin{equation}
    \dot{\varphi} = 0,
\end{equation}
\begin{equation}
    \dot{p} = 0,
\end{equation}
\begin{equation}
    \mathcal{D}_{10}(\kappa^2 pp' -\varphi' (e^\varphi \Lambda_0 -\varphi'')) = 0,
\end{equation}
\begin{equation}
    \mathcal{D}_{00}(\kappa^2 pp' -\varphi' (e^\varphi \Lambda_0 -\varphi'')) = 0,
\end{equation}
\begin{equation}
    \dot{p}_\perp = 2\mathcal{D}_{00},
\end{equation}
and
\begin{equation}
    \dot{p}_1 = 2\mathcal{D}_{10}.
\end{equation}
It is easy to see that two of the equations are linear combinations and simply form the condition
\begin{equation}
    \kappa^2 pp' -\varphi' (e^\varphi \Lambda_0 -\varphi'') = 0.
    \label{thing1}
\end{equation}

This condition recovers the canonical commutation relations. We now reduce to the constraint surface and find as a consistency  condition $\mathcal{D}_{00} \rightarrow 0$ and  $\mathcal{D}_{10} \rightarrow 0$. This fully recovers the Hamiltonian constraint $\mathcal{H}_c \approx 0$. Furthermore the secondary constraints become
\begin{equation}
         p\varphi' - 2p'  \approx 0,
         \label{thing2}
\end{equation}
and
\begin{equation}
        \frac{\kappa}{2}p^2 + \frac{1}{\kappa}\varphi'' -\frac{1}{2\kappa}\varphi'^2 + \frac{\Lambda_{0}}{\kappa}e^\varphi - \mathcal{D}_{11}
         \approx 0.
         \label{thing3}
\end{equation}
Thus in the proper-time gauge on the constraint surface the theory is described entirely by $\varphi, p$ and $\mathcal{D}_{11}$. Furthermore Eq.[\ref{thing1}], Eq.[\ref{thing2}], and Eq.[\ref{thing3}] fully define the system and can be solved for forms of the metric and its dependence on the diffeomorphism field. The momentum field is found to be
\begin{equation}
    p = \pm \frac{\sqrt{2}\sqrt{-b+b\tanh{[\tfrac{1}{2}\left(\pm \sqrt{2b}x \mp \sqrt{2b}c \right)]}^2}}{\sqrt{2e^a \Lambda_0 - \kappa^2}}
\end{equation}
and the conformal factor 
\begin{equation}
    \varphi = a + 2\log{p}.
\end{equation}
The constants $b$ and $c$ are determined from the diffeomorphism field as
\begin{equation}
    \mathcal{D}_{11} = \frac{b\left(5+\cosh[\sqrt{2b}(c-x)]\text{sech}[\tfrac{\sqrt{b}(c-x)}{\sqrt{2}}]^2 \right)}{2\kappa}.
\end{equation}
However, it is of note that neither the constant $a$ or the cosmological constant appear in this expression. Whereas the Ricci scalar for the ADM formalism in the proper-time gauge on this constraint surface becomes
\begin{equation}
    R = \frac{\kappa^2}{2}e^{-a}-\Lambda_0.
\end{equation}
The diffeomorphism field determines the specific form of the conformal factor, and therefore the metric, but it will have no effect on the overall curvature.

\section{Dynamical Diffeomorphism Field \label{III}}
We now investigate the dynamics of the diffeomorphism field in the Minkowski gauge by the inclusion of the projective Gauss-Bonnet terms. Both the Effective Polyakov action and the Einstein-Hilbert action are trivial in the two-dimensional Minkowski gauge. Thus, this theory is completely described by the projective Gauss-Bonnet action plus the trace of the diffeomorphism field. The action for this choice of metric becomes
\begin{widetext}
\begin{multline}
    S = \int{d^2x\left(\frac{\mathcal{D}_{11}-\mathcal{D}_{00}+\Lambda_0}{2\kappa}-J_0c(4\mathcal{D}_{10}^2-2\mathcal{D}_{00}\mathcal{D}_{11}-\mathcal{D}_{00}^2-\mathcal{D}_{11}^2)\right.}\\\left.+J_0c\lambda_0^2 (2(\partial_t \mathcal{D}_{10})^2-2(\partial_t \mathcal{D}_{11})^2-4(\partial_t \mathcal{D}_{10})(\partial_x \mathcal{D}_{00})+2(\partial_x \mathcal{D}_{00})^2 +4(\partial_t \mathcal{D}_{11})(\partial_x \mathcal{D}_{10})-2(\partial_x \mathcal{D}_{10})^2
    \right)
\label{eq:wideeq}
\end{multline}
\end{widetext}
At this stage, the coupling constant, $\kappa$, has not been associated with Newton's gravitational constant.
The canonical momenta for the diffeomorphism field is found to be
\begin{equation}
    p^{11} \equiv \frac{\delta S}{\delta \mathcal{D}_{11}} = 4J_0c\lambda_0^2((\partial_x \mathcal{D}_{10})-(\partial_t \mathcal{D}_{11})),
\end{equation}
\begin{equation}
    p^{10} \equiv \frac{\delta S}{\delta \mathcal{D}_{10}} = 4J_0c\lambda_0^2((\partial_t \mathcal{D}_{10})-(\partial_x \mathcal{D}_{00})),
\end{equation}
and
\begin{equation}
    p^{00} \equiv \frac{\delta S}{\delta \mathcal{D}_{00}} = 0.
    \label{minkprimcon}
\end{equation}
As $\mathcal{D}_{00}$ has no conjugate momenta, we promote Eq.[\ref{minkprimcon}] to the primary constraint $\phi \equiv p^{00} \approx 0$. Forming a first Hamiltonian and using the Poisson brackets over the naïve phase-space to find the secondary constraint
\begin{equation}
    \chi \equiv 2J_0c(\mathcal{D}_{00}+\mathcal{D}_{11})+\partial_x p^{10} -\frac{1}{2\kappa}.
\end{equation}
No further constraints are found in this theory. The two constraints are second-class with the constraint matrix
\begin{equation}
    \mathcal{M} = \begin{pmatrix}
        0 && -2J_0c\\
        2J_0c && 0
    \end{pmatrix}\delta(x-x').
\end{equation}
We can use the secondary constraint to write the Hamiltonian independent of $\mathcal{D}_{00}$. As $p^{00}$ is the primary constraint, the theory exhibits a gauge freedom of the $\mathcal{D}_{00}$ term of the diffeomorphism field. The constrained Hamiltonian density becomes
\begin{multline}
    \mathcal{H}_c = \frac{(p^{10})^2-(p^{11})^2}{8J_0c\lambda_0^2} +\frac{\kappa(\partial_x p^{10})^2-(\partial_xp^{10})}{4J_0 c\kappa}\\ +(\partial_x\mathcal{D}_{10})p^{11}-(\partial_x\mathcal{D}_{11})p^{10} +4J_0c(\mathcal{D}_{10})^2\\
    +\frac{1}{16J_0c\kappa^2} - \frac{\Lambda_0}{2\kappa}
\end{multline}

The dynamics of the diffeomorphism field found from the constrained Hamiltonian density and the Dirac brackets are
\begin{equation}
    \partial_t \mathcal{D}_{11} = \partial_x\mathcal{D}_{10}-\frac{p^{11}}{4J_0c\lambda_0^2},
\end{equation}
\begin{equation}
    \partial_t \mathcal{D}_{10} = -\partial_x\mathcal{D}_{11}-\frac{\partial_x^{\;2}p^{10}}{2J_0c}+\frac{p^{10}}{4J_0c\lambda_0^2},
\end{equation}
\begin{equation}
    \partial_t p^{11} = \frac{1}{\kappa}-\partial_x p^{10},
\end{equation}
and
\begin{equation}
    \partial_t p^{10} = \partial_x p^{11} - 8J_0c \mathcal{D}_{10}.
\end{equation}
From the Dirac brackets, it can be shown that the constrained phase-space fields exhibit the canonical commutation relations
\begin{equation}
    \{\mathcal{D}_{ab},p^{cd}\} = \delta_a^{c}\delta_b^{d}\delta(x-x').
\end{equation}
To investigate possible gauge transformations generated from the constraints, we construct the charges
\begin{equation}
    Q_1 = \int{dx V(x)\phi},\text{ and } Q_2 = \int{dx V(x)\chi}.
\end{equation}
From these charges, the Dirac bracket relations are found
\begin{equation}
    \{Q_1,\mathcal{D}_{ab}\} = -V\delta^0_a\delta^0_b, \;\{Q_1,p^{ab}\} = 0,
\end{equation}
and
\begin{multline}
    \{Q_2,\mathcal{D}_{ab}\} = \delta^1_{a}\delta^0_{b}\partial_x V,\\ \;\{Q_2,p^{ab}\} = 2J_0cV\left(\delta^a_0\delta^b_0 + \delta^a_1 \delta^b_1 \right).
\end{multline}

These equations of motion can generally be solved by the wave solutions
\begin{multline}
    \mathcal{D}_{11} = J_2 \cos(kx-\omega t)+J_1 \cos(kx+\omega t)\\ + G_2 \sin(kx-\omega t)+G_1 \sin(kx+\omega t) + \frac{x^2}{8J_0 c\lambda_0^2 \kappa},
\end{multline}
\begin{multline}
    \mathcal{D}_{10} = \frac{(\omega^2 + k^2(2\lambda_0^2 \omega^2))}{2k\omega(1+k^2\lambda_0^2)}\left(J_1 \cos(kx+\omega t)\right.\\\left.-J_2 \cos(kx-\omega t) + G_1 \sin(kx+\omega t)\right.\\\left.-G_2 \sin(kx-\omega t)\right),
\end{multline}
\begin{multline}
    p^{11} = \frac{2J_0c\lambda_0^2(\omega^2 + k^2)}{\omega(1+k^2\lambda_0^2)}\left(J_1 \sin(kx+\omega t)\right.\\\left.-J_2 \sin(kx-\omega t) - G_1 \cos(kx+\omega t)\right.\\\left.+G_2 \cos(kx-\omega t)\right),
\end{multline}
\begin{multline}
    p^{10} = \frac{2J_0c\lambda_0^2(k^2+\omega^2)}{k+k^3\lambda_0^2}\left(J_1 \sin(kx+\omega t) \right.\\\left. -J_2 \sin(kx-\omega t) +G_1 \cos(kx+\omega t) \right.\\\left. - G_2 \cos(kx-\omega t) 
    \right) + \frac{x}{\kappa}.
\end{multline}
With the speed of light wave propagation solution or the gravitational wave dispersion relation
\begin{equation}
    k = \pm\omega,\text{ and } k = \pm \frac{\sqrt{-2+\omega^2\lambda_0^2}}{\lambda_0}.
\end{equation}
Here $J_1, J_2, G_1, \text{ and } G_2$
are unknown scaling constants.
\subsection{Decomposition of the Diffeomorphism Field}
Investigated in \cite{Brensinger:2019mnx}, the diffeomorphism field is decomposed into a traceless portion,  $(g^{ab}W_{ab} = 0)$, and a scalar field, $\varphi$, as
\begin{equation}
    \mathcal{D}_{ab} = W_{ab} + \frac{M_\varphi c}{2J_0}g_{ab}\varphi + 2g_{ab}\Lambda_0.
    \label{diffdecomp}
\end{equation}
In the previous discussion the decomposition work is done in general and four dimensions; as in the rest of this paper, two-dimensions will be investigated. The field equations for the projective Gauss-Bonnet action with the diffeomorphism field decomposition in two-dimensions finds
\begin{multline}
    \mathcal{W}_{10} = 4J_0c\lambda_0^2\left(\partial_x^{\;2}W_{10} - \partial_t^{\;2}W_{10}
    \right)\\- 8J_0cW_{10}-4M_\varphi c^2 \lambda_0^2\partial_x\partial_t \varphi,
\end{multline}
\begin{multline}
    \mathcal{W}_{00} = 4J_0c\lambda_0^2\left(\partial_t^{\;2}W_{00} - \partial_x^{\;2}W_{00}
    \right)\\+ 8J_0cW_{00}+2M_\varphi c^2 \lambda_0^2\left(\partial_x^{\;2} \varphi + \partial_t^{\;2} \varphi\right),
\end{multline}
and
\begin{multline}
    \mathcal{\varPhi} = \frac{M_{\varphi}^2c^3\lambda_0^2}{J_0}\left(\partial_t^{\;2} \varphi + \partial_x^{\;2} \varphi\right)\\
    +2M_\varphi c^2 \lambda_0^2\left(\partial_t^{\;2}W_{00} + \partial_x^{\;2}W_{00} - \partial_x\partial_t W_{10}
    \right)\\+ \frac{M_\varphi c}{2J_0 \kappa}.
\end{multline}
Using the field equations on-shell and investigating the $W_{ab} \rightarrow 0$ choice finds
\begin{equation}
    \mathcal{W}_{10} \rightarrow 0,\;\; \mathcal{W}_{00} \rightarrow 0,
\end{equation}
and
\begin{equation}
    \mathcal{\varPhi} \rightarrow \frac{M_\varphi c}{2J_0 \kappa}.
\end{equation}
As the remaining condition on $\mathcal{\varPhi}$ takes the mass of the trace portion of the diffeomorphism field to be zero, it can be seen that on shell, for the traceless portion of the diffeomorphism field to vanish, the field equations constrain the diffeomorphism field such that $\mathcal{D}_{ab} \rightarrow 0$. Thus, imposing $W_{ab}$ to vanish will not be done throughout the rest of this paper. In \cite{InfProj} a Lagrange multiplier term is included by hand in the action to avoid this issue. As we will not be taking $W_{ab} = 0$, and to avoid confusion with Lagrange multiplier arising from the constraint analysis this Lagrange multiplier will not be used here. The Hamiltonian previously constructed in \cite{Brensinger:2019mnx} utilized vanishing $W_{ab}$ to investigate the dynamical portions of $\varphi$. The Hamiltonian constructed here will preserve the full decomposed diffeomorphism field and investigate the full projective Gauss-Bonnet action.

Special care should be used in constructing the decomposition Eq.[\ref{diffdecomp}] as the diffeomorphism field is a portion of a connection it is not tonsorial while the terms of the decomposition are. Using the diffeomorphism field and the metric compatible connection we can construct a tensor object $\mathcal{P}_{ab}$ as
\begin{equation}
    \mathcal{P}_{ab} = \mathcal{D}_{ab}- \partial_a \alpha_b + \Gamma^{a}_{\;\;bc}\alpha_c + \alpha_a\alpha_b.
\end{equation}
Where alpha is the one-form found in section $I$. It can be seen that for a Minkowski metric we have $\alpha_a = 0$. Thus it is possible to form such a decomposition of the diffeomorphism field for this choice of metric. Further discussion of the covariant and projectively invariant gravitational theory can be found in \cite{CovAndProj}.

The canonical momenta for the diffeomorphism field are found to be
\begin{equation}
    p^{00} = 4J_0c\lambda_0^2(\partial_x W_{10}-\partial_t W_{00}) -2M_\varphi c^2 \lambda_0^2 \partial_t \varphi,
\end{equation}
\begin{equation}
    p^{10} = 4J_0c\lambda_0^2(\partial_t W_{10}-\partial_t W_{00}) +2M_\varphi c^2 \lambda_0^2 \partial_x \varphi,
\end{equation}
and for $\varphi$
\begin{equation}
    p = 2M_\varphi c^2 \lambda_0^2 (\partial_x W_{10} - \partial_t W_{00}) - \frac{M_\varphi^2 c^3 \lambda_0^2}{J_0}\partial_t \varphi.
\end{equation}
The $\varphi$ canonical momenta can be rewritten using the other two to find
\begin{equation}
    p = \frac{M_\varphi c}{2J_0}p^{00}.
\end{equation}
This momenta is to be promoted to a constraint as
\begin{equation}
    \phi \equiv p - \frac{M_\varphi c}{2J_0}p^{00} \approx 0.
\end{equation}
Constructing a first Hamiltonian, we find the secondary constraint
\begin{equation}
    \chi \equiv \frac{M_\varphi c}{2J_0 \kappa}- 4M_\varphi c^2 W_{00} - \frac{M_\varphi c}{J_0}\partial_x p^{10}.
\end{equation}
There are no further constraints in this theory. The found constraints are found to be second-class with the constraint matrix
\begin{equation}
    \mathcal{M} = \begin{pmatrix}
        0 && -2J_0^{-1}M_\varphi^2 c^3\\
       2J_0^{-1}M_\varphi^2 c^3 && 0\\
    \end{pmatrix}\delta(x-x').
\end{equation}
Utilizing Dirac brackets, we are now free to constrain the Hamiltonian density. This finds
\begin{multline}
    \mathcal{H}_c = \frac{(p^{10})^2}{8J_0 c\lambda_0^2} - \frac{J_0 }{2M_\varphi^2 c^3 \lambda_0^2}p^2 - \frac{M_\varphi c }{4J_0\kappa}\varphi \\+ \frac{2J_0}{M_\varphi c}p\partial_x W_{10} + 4J_0c(W_{10}^2 - W_{00}^2)\\ - 2M_\varphi c^2 W_{00}\varphi + p^{10}\partial_xW_{00} -\frac{2\Lambda_0}{\kappa}.
\end{multline}
However, as we are taking the constraints to be strongly zero, we are free to add the constraints with an arbitrary factor to the theory. For a special choice of the Lagrange multiplier, we can cast the constrained Hamiltonian into the form
\begin{equation}
    \mathcal{H}'_c = \frac{(p^{10})^2 - (p^{00})^2}{8J_0 c \lambda_0^2}+p^{10}\partial_x W_{00} + p^{00}\partial_x W_{10}.
\end{equation}
This form of the Hamiltonian is independent of $\varphi$ and its momenta. As $\varphi$ does not appear in the constraints, it is fully defined from the consistency conditions of the Lagrange multipliers. The arbitrariness of $\varphi$ complicates solving the equations of motion, and thus the first form of the constrained Hamiltonian will be used to find field solutions to Hamilton's equations.

The dynamics of the diffeomorphism field found from the Dirac brackets are
\begin{equation}
    \partial_t \varphi = \frac{J_0}{M_\varphi^2 c^3}\partial_x^{\;2}p - \frac{J_0}{M_\varphi^2 c^3 \lambda_0^2}p - \frac{2J_0}{M_\varphi c}\partial_x W_{10},
\end{equation}
\begin{equation}
    \partial_t W_{00} = 2\partial_x W_{10} - \frac{\partial_x^{\;2}p}{2M_\varphi c},
\end{equation}
and
\begin{equation}
    \partial_t W_{10} = \frac{p^{10}}{4J_0 c\lambda_0^2} + \partial_x W_{00}.
\end{equation}
The dynamics of the momentum fields are found as 
\begin{equation}
    \partial_t p = \frac{M_\varphi c}{4J_0 \kappa} + 2M_\varphi c W_{00},
\end{equation}
\begin{equation}
    \partial_t p^{00} = \frac{1}{2\kappa} + 4J_0 c W_{00},
\end{equation}
and
\begin{equation}
    \partial_t p^{10} = \frac{2J_0}{M_\varphi c}\partial_x p - 8J_0 c W_{10}.
\end{equation}
The Dirac brackets of the traceless portion of the diffeomorphism field with its canonical momenta finds
\begin{equation}
    \{W_{ab},p^{cd}\} = \delta_c^c\delta_b^d \delta(x-x'),
\end{equation}
and for the scalar field
\begin{equation}
    \{\varphi,p\} = \delta(x-x').
\end{equation}
All other combinations of Dirac brackets of the phase-space fields are zero. Solutions to the equations of motion that satisfy the constraints are
\begin{multline}
    W_{00} = \frac{\omega}{4J_0 c}(Z_1 \cos(kx+\omega t)-Z_2 \cos(kx-\omega t)\\
    +P_2 \sin(kx-\omega t) - P_1 \sin(kx-t\omega)) + \frac{1}{8J_0 c\kappa},
\end{multline}
\begin{multline}
    W_{10} = \frac{k^2 + \omega^2}{8J_0 c \kappa}(Z_2 \cos(kx-\omega t) + Z_1 \cos(kx+\omega t)\\ - P_1 \sin(kx+t\omega) - P_2 \sin(kx-t\omega))\\+\frac{1}{4\sqrt{2}J_0 c\lambda_0}(b_1 \sin(\tfrac{\sqrt{2}t}{\lambda_0})-b_2 \cos(\tfrac{\sqrt{2}t}{\lambda_0}))\\
    +\frac{1}{2\sqrt{2}M_\varphi c^2 \lambda_0}((a_1-a_2)\cosh(\tfrac{\sqrt{2}x}{\lambda_0})\\+(a_1+a_2)\sinh(\tfrac{\sqrt{2}x}{\lambda_0})),
\end{multline}
\begin{multline}
    \varphi = \frac{2+k^2\lambda_0^2 -\omega^2 \lambda_0^2 }{4M_\varphi c^2 \lambda_0^2 \omega}(Z_1 \cos(kx+\omega t)\\-Z_2 \cos(kx-\omega t)+P_2 \sin(kx-t\omega)\\-P_1 \sin(kx+t\omega)) -\frac{t^2}{4M_\varphi c^2 \lambda_0^2 \kappa}\\ + b_3 t + \varphi_1(x) + b_4,
\end{multline}
\begin{multline}
    p^{00} = P_2 \cos(kx-\omega t) + P_1 \cos(kx+\omega t)\\ +Z_2 \sin(kx-t\omega) + Z_1 \sin(kx+t\omega)\\
    +\frac{2(a_1+a_2)J_0}{M_\varphi c}\cosh(\tfrac{\sqrt{2}x}{\lambda_0})\\ + \frac{2(a_1-a_2)J_0}{M_\varphi c}\sinh(\tfrac{\sqrt{2}x}{\lambda_0})+\frac{t}{\kappa}- 2b_3 M_\varphi c^2\lambda_0^2,
\end{multline}

\begin{multline}
    p^{10} = \frac{\lambda_0^2(k^2+\omega^2)\omega}{2k}(P_1 \cos(kx+\omega t)\\-P_2 \cos(kx-\omega t)+Z_1 \sin(kx+t\omega)\\-Z_2 \sin(kx-t\omega))+b_1 \cos(\tfrac{\sqrt{2}t}{\lambda_0})+b_2 \sin(\tfrac{\sqrt{2}t}{\lambda_0}),
\end{multline}
and
\begin{multline}
    p = \frac{M_\varphi c}{2J_0}(P_1 \cos(kx+\omega t)+P_2 \cos(kx-\omega t)\\+Z_1 \sin(kx+t\omega)+Z_2 \sin(kx-t\omega))\\+(a_1-a_2)\\ \sinh(\tfrac{\sqrt{2}x}{\lambda_0})+(a_1+a_2)\cosh(\tfrac{\sqrt{2}x}{\lambda_0})\\ +\frac{M_\varphi c}{2J_0 \kappa}(t - 2b_3 M_\varphi c^2 \lambda_0^2 \kappa).
\end{multline}
With the gravitational wave dispersion relation
\begin{equation}
    k = \pm \frac{\sqrt{-2+\omega^2\lambda_0^2}}{\lambda_0}.
\end{equation}
One will also find $\omega = 0$ solutions; however, this leads to indeterminate forms of the diffeomorphism field and is not considered a physical solution. In the solutions $Z_1, Z_2, P_1, P_2, a_1, a_2, b_1,b_2, b_3,\text{ and } b_4$ are unknown scaling factors, while $\varphi_1(x)$ is an unknown function dependent only on the $x$ coordinate. As $\varphi$ only appears in its own equation of motion, any x-dependent contribution can be added without changing the physics.

\section{Fully Dynamical Theory}
We will now investigate the effective Polyakov action coupled to the projective Gauss-Bonnet terms as a theory of dynamical diffeomorphism field and a dynamical metric. To avoid possible instabilities we use the ADM formalism for the metric and the P-tensor form of the diffeomorphism field. The action will not be presented here in its entirety as it lengthy, However, it can easily be found from Eq.[\ref{ptensor}], Eq.[\ref{PGBaction}] and Eq.[\ref{ADMEPAaction}]. The action is now
\begin{equation}
    S = F + S_{PGB}.
\end{equation}

As $\mathcal{P}_{ab}$ is reserved for the P-tensor we will denote the conjugate momenta as $E^{ab}$. We find these momenta as
\begin{equation}
    E^{ab} = \frac{\delta S}{\delta \dot{\mathcal{P}}_{ab}},
\end{equation}
while the momenta conjugate to the metric fields are
\begin{equation}
    \pi \equiv \frac{\delta F}{\delta \dot{\varphi}}, \pi_\perp  \equiv \frac{\delta S}{\delta \dot{\eta}^\perp}, \text{ and } \pi_1  \equiv \frac{\delta S}{\delta \dot{\eta}^1}.
\end{equation}
From these the three primary constraints are found
\begin{equation}
    \phi_\perp \equiv \pi_\perp +\frac{E^{10}\mathcal{P}_{10}}{\eta^\perp} - \frac{E^{10}\mathcal{P}_{11}\eta^1}{\eta^\perp} \approx 0,
\end{equation}
\begin{equation}
    \phi_1 \equiv \pi_1 +E^{10}\mathcal{P}_{11} \approx 0,
\end{equation}
and
\begin{equation}
    \phi  \equiv E^{00} \approx 0. 
\end{equation}

Constructing the first Hamiltonian density as $\mathcal{H}_1$, as before, reveals secondary constraints 
\begin{multline}
    \chi_\perp \equiv \{\phi_\perp,H_1\} \approx 0,\\ \; \chi_1 \equiv \{\phi_\perp,H_1\} \approx 0,\\\; \text{and}\; \chi \equiv \{\phi,H_1\} \approx 0.
\end{multline}
No further constraints are found, and both the primary and secondary constraints are second class. It is straightforward to show that the total Hamiltonian vanishes up to total derivatives and finds
\begin{equation}
    \mathcal{H} = V_{\perp}\phi_{\perp}+V_1 \phi_1 + V \phi + U_{\perp} \chi_\perp + U_{1}\chi_1 + U\chi \approx 0.
\end{equation}
Where $V_{\perp},\; V_1,\; V,\;U_{\perp},\; U_{1},\; \text{and}\; U$ are Lagrange multipliers. Thus, the Hamiltonian constraint is recovered for the fully dynamical theory. 

In the proper-time gauge with $\eta^\perp \rightarrow 1$ and $\eta^1 \rightarrow 0$ the primary constraints are reduced to
\begin{equation}
    \phi_{\perp} = E^{10}\mathcal{P}_{10} \approx 0,
\end{equation}
\begin{equation}
    \phi_1 = E^{10}\mathcal{P}_{11} \approx 0,
\end{equation}
and
\begin{equation}
    \phi = E^{00}\approx 0.
\end{equation}
Interpretation of the $\phi$ constraint is straightforward as necessitating $E^{00} \rightarrow 0$ demonstrating a gauge freedom with $\mathcal{P}_{00}$. However, the first two constraints can be satisfied by either $\mathcal{P}_{10} \rightarrow 0$ and $\mathcal{P}_{11} \rightarrow 0$ or by simply $E^{10} \rightarrow 0$. As $\mathcal{P}_{00}$ is previously established to have a gauge freedom, choosing the first option renders the P-tensor non-contributing to the remainder of the constraint equations. We will thus investigate the $E^{10} \rightarrow 0$ option; furthermore this can naturally be achieved from the action by having $\mathcal{P}_{10} \rightarrow 0$. The secondary constraints are now found to be
\begin{multline}
    \chi_\perp = \frac{2\partial_x^{\;2} \varphi-\tfrac{1}{2}(\partial_x \varphi)^2+e^{\varphi}\Lambda}{k}+\mathcal{P}_{11}\\+ J_0 c e^{\varphi}(\mathcal{P}_{11})^2-\frac{1}{2}k\pi^2 -\frac{1}{8}k(E^{11})^2(\mathcal{P}_{11})^2\\
    -\frac{1}{2}kE^{11}\mathcal{P}_{11}\pi +\frac{e^{2\varphi}(E^{11})^2}{8J_0 c \lambda_0^2} \approx 0,
\end{multline}
\begin{multline}
    \chi_1 = 2\mathcal{P}_{11}\partial_x E^{11}+E^{11}\partial_x \mathcal{P}_{11}\\ + 2\partial_x \pi -\pi \partial_x \varphi \approx 0,
\end{multline}
and
\begin{equation}
    \chi = 2J_0 ce^{-\varphi} \mathcal{P}_{11}+\frac{1}{4}k(E^{11})^2\mathcal{P}_{11}+\frac{1}{2}kE^{11}\pi - 1 \approx 0.
\end{equation}

The remaining secondary constraints do not completely define the four remaining degrees of freedom; as such future work will attempt to find general solutions that satisfy the constraint surface.

\section{Conclusion}
Treating the diffeomorphism field as a naturally occurring geometric contribution greatly modifies the canonical quantization of the effective Polyakov action. In both the dynamical light-cone gauge and the ADM formalism, we recover the vanishing Hamiltonians when the diffeomorphism field is a background field. In the dynamical light-cone gauge, we are able to solve the constraint using a Fourier mode expansion of the metric fields. The quantum operators corresponding to the metric are defined from the expectation value of the diffeomorphism field value in the form of a number operator with a $\kappa$ multiple. As a model where the diffeomorphism fields' value comes from a gravitational theory with a Minkowski metric, we can utilize the dispersion relation and the speed of light solutions found in section $\ref{III}$. For both cases, we find the same expectation value, no transition amplitudes, and successfully quantize the theory.

In the ADM formalism, the Hamiltonian constraint is written in terms of the diffeomorphism field. Reduction to the constraint surface of Hamilton's equations of motion reveal consistency conditions on the diffeomorphism, which fully recover the vanishing Hamiltonian. It is also found that the canonical commutation relations have a non-trivial form. Investigating the proper-time gauge finds simpler forms of the constraints, as well as two of the equations of motion being linear multiples of one another, required to vanish. These three equations fully define the system and can be solved for exact forms of the conformal field and its momentum. Furthermore, how the diffeomorphism field determines the unknown constants is found. However, it can be shown that of the four constants in the conformal field, the diffeomorphism field only defines two, while the Ricci scalar only depends on the other two (one of which being the cosmological constant).

Although appearing to take different forms in the ADM formalism versus the dynamical light-cone gauge, as these are gauge choices, the two forms of the quantum theory must be reconciled. This is only required on the constraint surface, and two gauge-independent results are demonstrated. First, the vanishing of the Hamiltonian is shown regardless of the gauge. As the Hamiltonian offers a natural mechanism for the evolution of the quantum state, the vanishing of the Hamiltonian must be recovered regardless of the choice of metric. The second gauge-independent result is the surviving degrees of freedom on the constraint surface. In the light-cone gauge, two components of the diffeomorphism field are found to be non-contributing to the action. In the ADM formalism all three degrees of freedom arising from the diffeomorphism field contribute. However, it is found that by reduction to the constraint surface, two degrees of freedom vanish. We thus recover the same number of degrees of freedom of the diffeomorphism field in all gauges investigated. Furthermore, as the diffeomorphism field is held as a background contribution to the theory, we find in both the dynamical light-cone gauge and the proper time gauge that on the constraint surface, only one component of the background field couples to the system.

Investigating the dynamical diffeomorphism field with a background Minkowski metric finds that the theory is fully built from the projective Gauss-Bonnet action. We find that both $\mathcal{D}_{00}$ and its canonical momenta $p^{00}$ are defined through secondary and primary constraints. On the constraint surface, we are able to write the theory independent of these fields. This reveals a gauge freedom in the $\mathcal{D}_{00}$ component of the diffeomorphism field. Classically, we are able to solve the equations of motion of the constrained diffeomorphism field using wave forms subject to a speed of light wave propagation or dispersion relation condition. We have also shown that for the decomposition of the diffeomorphism field into traceless and trace components, choosing the traceless component to vanish requires the full vanishing of the diffeomorphism field at the field equation level. We recover the dispersion relation condition for this decomposition.

We have thus been able to show that the diffeomorphism field has a non-trivial contribution to the vanishing Hamiltonian constraints. As a background field, although still recovering these constraints, the diffeomorphism field fully determines the quantum state of the metric. While the dynamical diffeomorphisms' non-scale invariant projective Gauss-Bonnet action removes this constraint structure and allows dynamics of the system. Utilizing the P-tensor formalism of the diffeomorphism allows us to write the theory once again in scale invariant form. The constraint analysis of the fully dynamical theory in this form recovers Hamiltonian constraint. Within the proper time gauge; reduction to the constraint surface reduces the phase-space to a single contribution from the diffeomorphism field and the metric.

Finally, the completely dynamical theory is investigated in the ADM formalism and utilizing the P-tensor form of the diffeomorphism field. It is demonstrated that the for the fully dynamical theory, the Hamiltonian constraint can still be recovered.

\begin{acknowledgments}
The authors would like to thank Kory Stiffler, Nicholas Harshman, and the other members of the Nuclear and Particle Theory Groups at the University of Iowa for helpful discussions.
\end{acknowledgments}

\bibliographystyle{apsrev4-1} 

\end{document}